\documentclass[twocolumn,showpacs,superscriptaddress,preprintnumbers,amsmath,amssymb]{revtex4}

\usepackage{graphicx}
\usepackage{dcolumn}
\usepackage{bm}
\usepackage{epsfig}
\usepackage{epstopdf}
\usepackage{amsmath}
\usepackage{amsfonts}

\newcommand{\be}{\begin{equation}}
\newcommand{\ee}{\end{equation}}
\newcommand{\bea}{\begin{eqnarray}}
\newcommand{\eea}{\end{eqnarray}}

\begin{document}

\title{Controllability on relaxation-free
  subspaces: On the relationship between adiabatic population transfer and
  optimal control}
\author{Haidong Yuan}
\email{haidong.yuan@gmail.com}
\affiliation{Research Laboratory of Electronics,
Massachusetts Institute of Technology, Cambridge, MA 02139}
\author{Christiane P. Koch}
\affiliation{Institut f\"ur Theoretische Physik, Freie Universit\"at
  Berlin, Arnimallee 14, 14195 Berlin, Germany}
\altaffiliation{now at: Institut f\"ur Physik, Universit\"at Kassel,
  Heinrich-Plett-Str. 40, 34132 Kassel, Germany}
\author{Peter Salamon}
\affiliation{Department of Mathematical Science,
San Diego State University, San Diego, California, 92182, USA}
\author{David J. Tannor}
\affiliation{Department of Chemical Physics,
Weizmann Institute of Science, 76100 Rehovot, Israel}
\date{\today}

\begin{abstract}
  We consider the optimal control problem of transferring population
  between states of a quantum system where the coupling proceeds only via
  intermediate states that are subject to decay.
  We pose the question whether it is generally possible to carry out
  this transfer. For a
  single intermediate decaying state, we recover the Stimulated Raman
  Adiabatic Passage (STIRAP) process which we identify as the global
  optimum in the limit of infinite control time.
  We also present analytical solutions for
  the case of transfer that has to proceed via two
  consecutive intermediate decaying states.
  We show that in this case, for finite power the optimal control does not
  approach perfect state transfer even in the infinite time limit.
  We generalize our findings to characterize the topologies of paths
  that can be achieved by coherent control under the assumption of
  finite power. If two or more consecutive states in an $N$-level
  chain are subject to decay, complete population transfer with
  finite-power controls is not possible.
\end{abstract}

\maketitle
\section{Introduction}
Stimulated Raman adiabatic passage (STIRAP) achieves coherent
population transfer in three-level atoms or molecules
despite the short lifetime
of the intermediate level \cite{BergmannRMP98}. The key is the
creation of a dark state produced by overlapping pump and Stokes pulses in
a counter-intuitive sequence. In the adiabatic limit, the intermediate
state then never gets populated. STIRAP was first demonstrated two
decades ago \cite{GaubatzJCP90} but it continues to enjoy great
popularity due to its simple, yet robust character
\cite{LangPRL08,NiSci08}.

\begin{figure}[bt]
\begin{center}
\includegraphics[scale=.5]{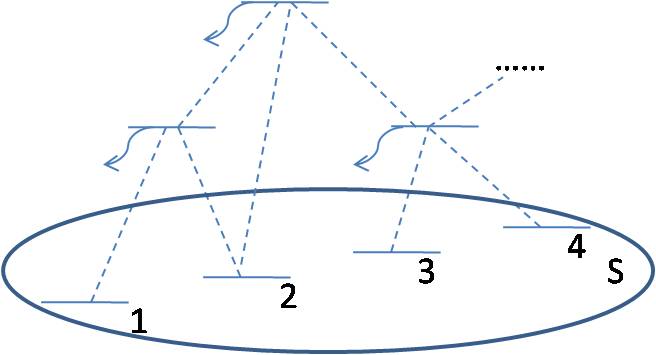}
\end{center}
\caption{(Color online)
  A general $N$-level system where the states in the subspace $S$ are
  connected through intermediate states outside of $S$.}
\label{fig:1}
\end{figure}

Inspired by this, we consider the general $N$-level system with a
subspace that is free of relaxation, as shown in
Fig.~\ref{fig:1}.
We consider the case that there is no direct coupling between states
in the relaxation-free subspace, but the states are coupled by
intermediate states that undergo relaxation.
The question we ask here is what kind of coupling
topology can ensure state-to-state controllability on the relaxation-free
subspace? We show that this question can be reduced to asking
what kind of coupling topology can
ensure unit efficiency of population transfer for any two eigenstates
in the relaxation-free subspace.

Our work is closely related to previous studies of STIRAP in
multi-level
chains~\cite{ShorePRA91,Tannor97,VitanovEPJD98,NakajimaPRA99}
which showed that under certain assumptions
the dark state condition can be generalized from the three-level to
the $N$-level case. In these studies, the decay from intermediate
levels was not explicitly taken into account. Here, we include
the dissipation. Moreover, we use an analytical formulation of optimal
control theory which
allows us to draw striking general conclusions about $N$-level
systems. In particular, we are able to show a relationship between
controllability and connectivity of decaying states in the chain:
A relaxation-free subspace is controllable on the pure state
space if and only if any two
eigenstates in the subspace can be connected by a path that never
visits two consecutive states that both suffer relaxation.
This coupling topology includes degenerate levels
provided that a generalized Morris-Shore transformation exists to
replace the coupled multi-level system
by a set of  two- and three-level systems and single dark
states~\cite{RangelovPRA06}.

The problem of controllability in a relaxation-free subspace
is closely related to fault tolerant quantum
computing and decoherence-free subspaces
in quantum information science.
Given that the Hamiltonian obeys a certain symmetry,
two or more physical qubits can be employed to encode one logical
qubit that is free of decoherence~\cite{LidarReviewDFS03}.
The condition for a dark state ensuring STIRAP turns out to be
equivalent to the condition for a decoherence-free
subspace to exist~\cite{LidarReviewDFS03}.
While in principle it is possible to construct
quantum gates that preserve the structure of the decoherence
free subspace \cite{MonzPRL09}, these gates are generally difficult
to implement in practice for the following reason:
the gate operations need to be carried out with controls that
act on the physical qubits and this introduces couplings to the
decohering subspaces \cite{FortunatoNJP02}.
This raises the question of whether losses can
still be avoided if the controls are chosen in an optimal way.
Previous work has discussed  whether optimal control can find
STIRAP-like solutions in $N$-level
chains~\cite{BandJCP94,WangJCP96,Tannor97,TannorPRA99,KisJMO02,Boscain02}. However,
none of these studies took the dissipation explicitly into account.

Our paper is organized as follows. In section~\ref{sec:three}, we
study the three-level system. Using optimal control theory we show that
a STIRAP-like process represents the globally optimal process
for the population transfer in the relaxation-free subspace.
This STIRAP-like process is the infinite time limit of analytic
solutions we find for the problem with finite time.
In section~\ref{sec:four}, we study a four-level chain
system where the two intermediate states are subject to decay.
We show that with limited pulse power, it is not possible to
achieve complete population transfer.
Section~\ref{sec:N} generalizes these results to $N$-level
chains. This generalization is used in section~\ref{sec:control}  to
state the conditions on state-to-state controllability on the relaxation-free
subspace. Section~\ref{sec:concl} concludes.

\section{Three-level system}\label{sec:three}
We consider a three-level $\Lambda$-system with states $|1\rangle$, $|2\rangle$,
$|3\rangle$ where $|2\rangle$ suffers relaxation loss with
rate $k$.  It is well known that population transfer from $|1\rangle$ to
$|3\rangle$ is possible by STIRAP without populating state $|2\rangle$, i.e.,
complete transfer is achieved in the adiabatic limit \cite{BergmannRMP98}.
In this section, we formulate the population transfer
as an optimal control problem, taking the dissipation explicitly into
account. Previous work has addressed this problem using a
numerical density matrix optimization \cite{HornungPRA02}. Here, we
formulate the problem analytically using the Hamilton-Jacobi-Bellman
method which has the additional advantage of allowing us to determine
the global optimum. We will show that STIRAP arises
naturally as the solution to the optimal control problem in the adiabatic
limit. By yielding an upper bound for the transfer efficiency in
finite time, this formulation also gives some insight into the three-level
system in the non-adiabatic regime.

The dynamics of the three-level system are described by the following
effective Schr\"odinger equation, 
\begin{equation}
  \frac{d}{dt}\left(\begin{array}{c}
      x'_1  \\
      x'_2  \\
      x'_3
    \end{array}\right)
  =-i\left(\begin{array}{ccc}
      0 & \Omega_p & 0 \\
      \Omega_p & -ik & \Omega_s \\
      0 & \Omega_s & 0
    \end{array}\right)
  \left(\begin{array}{c}
      x'_1  \\
      x'_2  \\
      x'_3
    \end{array}\right)
\end{equation}
where we assume that the detuning is zero.
$\Omega_p$ and $\Omega_s$ are half the Rabi frequencies of the pump and
Stokes pulses~\footnote{
  The factor $1/2$ arising from the rotating wave approximation has been
  absorbed into $\Omega_p$ and $\Omega_s$ for simplicity.
  }, and
$k$ is the decay rate of state $|2\rangle$. We want to
optimize the transfer of population from $|1\rangle$ to $|3\rangle$
within a given time $T$, i.e., to steer the system
from the initial state $|\phi(0)\rangle=(1,0,0)$ to the final state
$|\phi(T)\rangle$ such that $|x'_3(T)|$ is maximized.

We first make a change of variables, setting $x_1=x_1'$, $x_2=ix'_2$,
$x_3=-x_3'$. The dynamics becomes
\begin{equation}\label{eq:dyna}
\frac{d}{dt}\left(\begin{array}{c}
      x_1  \\
      x_2  \\
      x_3
\end{array}\right)
=\left(\begin{array}{ccc}
      0 & -\Omega_p & 0 \\
      \Omega_p & -k & -\Omega_s \\
      0 & \Omega_s & 0
\end{array}\right)
\left(\begin{array}{c}
      x_1  \\
      x_2  \\
      x_3
\end{array}\right)\,.
\end{equation}
Under these dynamics, if we start from the initial state $(1,0,0)$,
all the state variables will remain real, i.e.,  this change of
variables is motivated by  the structure of the Schr\"odinger equation
and the fact that we are looking for state-to-state control.
Initial states
such as $(e^{i\varphi},0,0)$ can be written as $e^{i\varphi}(1,0,0)$, and
since quantum operations act linearly on the states, these cases can
be reduced to the $(1,0,0)$ case. The goal is to
transfer population from $(1,0,0)$ to the final state such that $x_3(T)$ is
maximized under the controls $\Omega_p$ and $\Omega_s$\footnote{
  Note that this control assumes that our pulses are resonant with fixed
  carrier frequencies and phases.
}.

In a realistic setup,
both pump and Stokes pulses are limited in amplitude, but we will
relax this condition: we assume that $\Omega_s$ is
bounded in amplitude by $A$ while $\Omega_p$ is unbounded. This
assumption enables us to solve the problem analytically and yields an upper
bound on the transfer efficiency. Since the amplitude of the pulses is usually quite large
compared to the relaxation rate, these upper bounds are quite
tight. We will also show that in the
adiabatic limit, the condition of unbounded $\Omega_p$
can be relaxed and STIRAP-like
pulses arise naturally from the solution of the optimal control
problem.

\subsection{Optimal solution}
To find the optimal pulses, we make another change of variables,
setting
$r_1=\sqrt{x^2_1+x^2_2}$, $r_2=x_3$, and $\tan\theta=\frac{x_1}{x_2}$,
where $\theta\in[0,\frac{\pi}{2}]$. The dynamics for $(r_1,r_2)$ are
derived from Eq.(\ref{eq:dyna}),
\begin{equation}\label{eq:r1r2}
\frac{d}{dt}\left(\begin{array}{c}
      r_1  \\
      r_2  \\
\end{array}\right)
=\left(\begin{array}{cc}
      -k\cos^2\theta & -\Omega_s \cos\theta \\
      \Omega_s \cos \theta & 0 \\
\end{array}\right)
\left(\begin{array}{c}
      r_1  \\
      r_2  \\
\end{array}\right)\,.
\end{equation}
Note that $\Omega_p$ is now contained in $\theta$ and $r_1$
is related to the bright-state amplitude in STIRAP.
This change of variables is not intuitive but crucial for obtaining an
analytical solution. For equations of motion linear in the control,
the optimal control problem typically becomes singular and no
conditions to determine an analytical solution are obtained. After
this change of variables,
the equations of motion are non-linear in one
of the controls, $\cos\theta$. This will allow us to obtain a
non-vanishing condition determining the optimal solution when
applying Pontryagin's maximum principle.

Considering the physics of the problem we find
that $\Omega_s$ should take on the maximal
amplitude $A$ throughout the process:
We start from the initial state $(r_1,r_2)=(1,0)$ and want to maximize
$r_2(T)$. The population transfer between $r_1$ and $r_2$
depends on the rotation speed which is determined by
$\Omega_s\cos\theta$. At the same time, the population transfer is
compromised by $r_1$ undergoing decay. The effect of decay on
$r_1$ can be decreased by lowering the value of $\cos\theta$. In order to
keep the rotation speed between $r_1$ and $r_2$ constant,
$\Omega_s$ needs to be increased. For
a given rotation speed, the minimum value of  $\cos\theta$ and thus
the minimum effect of the decay is obtained for  $\Omega_s$ taking its
maximum value, $\Omega_s=A$.

We are now left with determining  $\Omega_p$, or, equivalently,
the angle $\theta$. Defining
$u=\cos\theta$, we rewrite the dynamics
\begin{equation}\label{eq:r}
\frac{d}{dt}\left(\begin{array}{c}
      r_1  \\
      r_2  \\
\end{array}\right)
=\left(\begin{array}{cc}
      -ku^2 & -A u \\
      A u & 0 \\
\end{array}\right)
\left(\begin{array}{c}
      r_1  \\
      r_2  \\
\end{array}\right).
\end{equation}
To determine the optimal control $u^*(t)$, we
use the principle of dynamic programming~\cite{Advances} and solve for
the maximum achievable value of $r_2(T)$ for all initial points
$(r_1 ,r_2)$. Starting from $(r_1 ,r_2)$, we denote the maximum
achievable value of $r_2$ by $V(r_1 ,r_2 ,t)$, also called the optimal
return function for the point $(r_1 ,r_2)$ at time t. Note that
for finite time problems, $T< \infty$, the optimal return function
has an explicit dependence on time, which has to satisfy the well known
Hamilton-Jacobi-Bellman equation,
\be
\label{eq:HJB}
\frac{\partial V}{\partial t}+\max_{u}H(u)=0 \,,
\ee
where
\be
\label{eq:Ham}
H(u)=
\left(\begin{array}{cc}
      \frac{\partial V}{\partial r_1} & \frac{\partial V}{\partial r_2} \\
\end{array}\right)
\left(\begin{array}{cc}
      -ku^2 & -A u \\
      A u & 0 \\
\end{array}\right)
\left(\begin{array}{c}
      r_1  \\
      r_2  \\
\end{array}\right)
\ee
is the Hamiltonian of the optimal control problem.

By solving the Hamilton-Jacobi-Bellman equation, we
obtain the optimal solution to the control problem. The
detailed derivation is presented in Appendix~\ref{sec:optimal}. Here
we just state the results we need for the subsequent analysis.
For control time $T$ longer than a critical time $T_M$ the optimal
control has two distinct phases:
\be \label{eq:uoptimal}
u^*(t) = \left\{ \begin{array}{lc}
\frac{1}{\sqrt{A^2(\tau^2-t^2)+2k(\tau-t)+1}} & \textrm{for $t\in[0,\tau]$}\\
1 & \textrm{for $t\in[\tau,T]$}\\
\end{array} \,,\right.
\ee
Expressions for $T_M$ and $\tau$ are given in Appendix~\ref{sec:optimal}.
The second stage  of this solution, $u^*(t)=1$, is a result of the
artificial choice of constraints in our formulation (bounded $\Omega_S$
and unbounded
$\Omega_P$).  But this need not concern us: as we show in
Section~\ref{subsec:stirap} the
second phase of the solution vanishes in the limit $T \to \infty$
(adiabatic limit), in which case we recover the STIRAP solution.

\subsection{Recovery of STIRAP}
\label{subsec:stirap}
We now show that in the limit $T\to\infty$,
our optimal pulse corresponds to STIRAP. The
optimal Rabi frequencies are derived from $r_1$, $r_2$ and $u^*$.
From Eq.(\ref{eq:dyna}), it follows that
\be
\frac{d}{dt}x_2=\Omega_px_1-kx_2-\Omega_sx_3 \,.
\ee
Substituting $x_2=r_1u$, $x_1=r_1\sqrt{1-u^2}$, and $x_3=r_2$, we obtain
\be
\label{eq:4omega}
\Omega_p=\frac{-ku^2r_1-\Omega_s ur_2+r_1\dot{u}+kr_1u+\Omega_sr_2}{r_1\sqrt{1-u^2}} \,.
\ee
When the control time goes to infinity, $T\rightarrow\infty$, the
switching time, $\tau$, also approaches $\infty$. This follows from
$T-\tau$ being smaller than the critical time $T_M$ as explained in
Appendix~\ref{sec:optimal}.
If  $\tau$ becomes very large, then
according to Eq.~(\ref{eq:uoptimal}), $u^*(t)$ is very small.
Therefore $x_2=r_1u^*\sim0$, i.e., the
level $|2\rangle$ is not populated. Substituting $u\sim0$ and $\dot{u}\sim0$ into
Eq. (\ref{eq:4omega}), we see that
\bea \label{eq:amplitudes}
\Omega_p =\frac{r_2}{r_1}\Omega_s
=\frac{x_3}{\sqrt{x_1^2+x_2^2}}\Omega_s
=\frac{x_3}{x_1}\Omega_s\,.
\eea
So at each time point, $x_2=0$,
$\frac{x_3}{x_1}=\frac{\Omega_p}{\Omega_s}$,
i.e., the system is in the state
$x_1|1\rangle+x_3|3\rangle$, where
$\frac{x_3}{x_1}=\frac{\Omega_p}{\Omega_s}$,  which corresponds to the
dark state in STIRAP.

\begin{figure}[tb]
  \centering
  \includegraphics[width=0.95\linewidth]{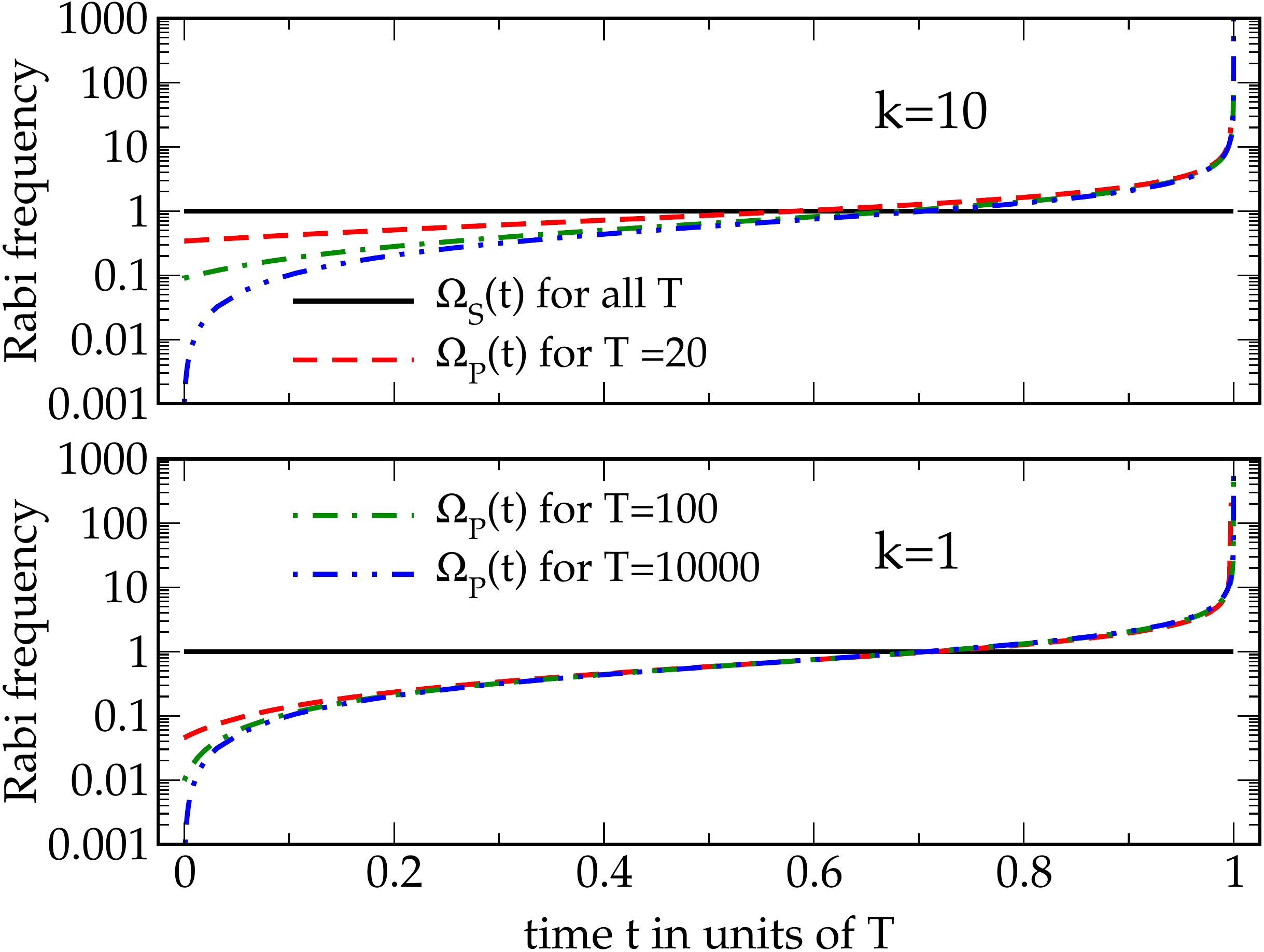}
  \caption{(Color online) Optimal half Rabi frequencies
    $\Omega_S(t)$ and $\Omega_P(t)$ for
    finite control times $T$ and
    relaxation rates $k$ (with the amplitude bound on
    $\Omega_S$ set to one, $A=1$).
    }
  \label{fig:fields}
\end{figure}

Note that in this limit, an upper bound can also be assumed for $\Omega_p$ in
addition to the one for $\Omega_s$. Since it is the ratio
$\Omega_p/\Omega_s$ that matters, one can simply lower $\Omega_s$ if
$\Omega_p$ exceeds its upper bound to
maintain the ratio $\Omega_p/\Omega_s$.
For example, one could assume the same bound $A$ on $\Omega_p$. In the
optimal solution shown in Fig.~\ref{fig:fields}, one would then
rescale time whenever $\Omega_p$ hits $A$. This obviously avoids the
shoot-off of  $\Omega_p$ to infinity visible in  Fig.~\ref{fig:fields}
at late times. Such a rescaling of amplitude corresponds to
changing the unit of time:  When $\Omega_p\geq A$,
a new time variable $d\tau=\frac{\Omega_p}{A}dt$ is defined.
In this new unit of time, the dynamics become
\begin{equation}\label{eq:rescale}
\aligned
\frac{d}{d\tau}\left(\begin{array}{c}
      x_1  \\
      x_2  \\
      x_3
\end{array}\right)
=&\frac{d}{dt}\left(\begin{array}{c}
      x_1  \\
      x_2  \\
      x_3
\end{array}\right)\frac{dt}{d\tau}\\
=&\left(\begin{array}{ccc}
      0 & -A & 0 \\
      A & -\frac{A}{\Omega_p} k & -\frac{A}{\Omega_p} \Omega_s\\
      0 &  \frac{A}{\Omega_p} \Omega_s& 0
\end{array}\right)
\left(\begin{array}{c}
      x_1  \\
      x_2  \\
      x_3
\end{array}\right).
\endaligned
\end{equation}
The relaxation rate $k$ does not affect the dynamics in the infinite
time limit, so $k$ and $\frac{A}{\Omega_p}k$ are effectively the same,
and the ratio of the Rabi
frequencies still satisfies the optimal condition in
Eq.~\eqref{eq:amplitudes}. Since the reparametrized time remains
infinite in the limit $T\rightarrow\infty$, optimality of
our solution is not affected by changing the unit of time.

One might wonder in this case
where the characteristic time delay between $\Omega_s$
and $\Omega_p$ is hidden in our solution. The point is that STIRAP is determined
by the overlap of the pulses. The rising part of the Stokes pulse when
the pump is zero and the falling part of the pump pulse when the
Stokes is zero do not affect the system dynamics. Our solution only
contains the crucial overlapping part, similar to the "shark-fin"
pulses discussed in Ref.~\cite{YatsenkoEPJD98}.
The Stokes pulse starts out
fairly flat at the upper bound and falls down as the pump pulse is
rising up to the bound. So the delay between the pulses in
standard STIRAP corresponds to the time the pump pulse takes to rise to the upper bound.
In our solution, a rising edge of the Stokes pulse and a falling edge
of the pump pulse could be added if one wishes to obtain a more
realistic pulse shape.

To summarize the similarities and differences with the conventional
STIRAP solution,  we drop the constraint on the bound of the pulse
amplitude $\Omega_p$ and obtain an analytical solution for the optimal
control (cf. Eqs.~\eqref{eq:uoptimal},\eqref{eq:4omega}) in finite
time. While in general the optimal pulse shape can be found only
numerically, for infinite time we obtain a completely analytical
solution (cf. Eq.~\eqref{eq:amplitudes}).
By solving the
Hamilton-Jacobi-Bellman equation, we have proven that this solution
is the global optimum. The rise of the Stokes pulse
and the fall of the pump pulse are missing. These portions  of the
pulses are irrelevant in the infinite time limit.
Thus, our analytical solution confirms that the essential feature of
STIRAP is the time-ordering of the pump and Stokes pulse where they
overlap.

\section{Four-level system}\label{sec:four}
\begin{figure}[bt]
\begin{center}
\includegraphics[scale=.5]{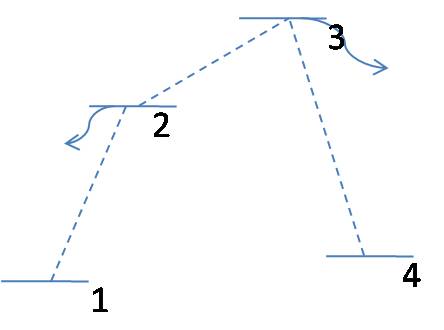}
\end{center}
\caption{(Color online) A four-level system: Population shall be transferred from
  $|1\rangle$ to $|4\rangle$ via the intermediate states $|2\rangle$
  and $|3\rangle$ which suffer relaxation.}
   \label{fig:3}
\end{figure}

In this section, we extend our method to a four-level chain system as shown
in Fig.~\ref{fig:3}. Again we explicitly include the decay from
the intermediate levels. We can thus demonstrate that the four-level
chain differs fundamentally from the
three-level system where STIRAP-like processes can transfer the
population fully in the adiabatic
limit. It turns out that in four-level systems with two decaying
intermediate states it is not possible to achieve complete transfer
with limited power even if we wait infinite time.

The dynamics of this system are described by the following
effective Schr\"odinger equation,
\begin{equation}
\frac{d}{dt}\left(\begin{array}{c}
      x'_1  \\
      x'_2  \\
      x'_3  \\
      x'_4
\end{array}\right)
=-i\left(\begin{array}{cccc}
      0 & \Omega_p & 0 & 0\\
      \Omega_p & -ik & \Omega_I & 0 \\
      0 & \Omega_I & -ik & \Omega_s\\
      0 & 0 & \Omega_s & 0
\end{array}\right)
\left(\begin{array}{c}
      x'_1  \\
      x'_2  \\
      x'_3  \\
      x'_4
\end{array}\right)\,,
\end{equation}
where $\Omega_p$ and $\Omega_s$ are the Rabi frequencies of
pump and Stokes pulses, $\Omega_I$
is the Rabi frequency of a pulse coupling $|2\rangle$ and $|3\rangle$,
and $k$ denotes the decay rate. As in the previous section, we assume
zero detunings.
Since we are only interested to find out whether there are schemes which
avoid populating the intermediate states, the exact value of
the relaxation rate is not important here. For simplicity, we assume
that the intermediate states suffer the same amount of relaxation. It is
straightforward to generalize to the cases with different decay
rates.

We want to find the optimal way to transfer population from
$|1\rangle$ to $|4\rangle$ within a given time $T$, i.e., the
optimal way of steering the system from the initial state
$|\phi(0)\rangle=(1,0,0,0)$ to the final state $|\phi(T)\rangle$ such
that $|x'_4(T)|$ is maximized.

Analogously to Section~\ref{sec:three}, we make a first change of variables,
letting $x_1=x_1'$, $x_2=ix'_2$,
$x_3=-x_3'$, $x_4=-ix_4'$. The dynamics become
\begin{equation}\label{eq:dyna4}
  \frac{d}{dt}\left(\begin{array}{c}
      x_1  \\
      x_2  \\
      x_3  \\
      x_4
    \end{array}\right)
  =\left(\begin{array}{cccc}
      0 & -\Omega_p & 0 & 0\\
      \Omega_p & -k & -\Omega_I & 0 \\
      0 & \Omega_I & -k & -\Omega_s \\
      0 & 0 & \Omega_s & 0
    \end{array}\right)
  \left(\begin{array}{c}
      x_1  \\
      x_2  \\
      x_3  \\
      x_4
\end{array}\right)\,.
\end{equation}
The state variables are now all real numbers if we start from the
initial state $(1,0,0,0)$. We want to transfer from $(1,0,0, 0)$
to the final state such that $x_4(T)$ is maximized under the controls
$\Omega_p$, $\Omega_I$ and $\Omega_s$ at given time $T$. We again
relax the control constraint, assuming that $\Omega_I$ is bounded in
amplitude by $A$, but $\Omega_p$ and $\Omega_s$ are not bounded. We
show that even with these relaxed control constraints, it is not
possible to achieve unit transfer efficiency.

To solve the problem, we make a second change of variables, letting
$r_1=\sqrt{x_1^2+x_2^2}$, $r_2=\sqrt{x_3^2+x_4^2}$,
$\tan\theta_1=\frac{x_1}{x_2}$, $\tan\theta_2=\frac{x_4}{x_3}$. The
dynamics of $(r_1,r_2)$ become
\begin{equation}
\frac{d}{dt}\left(\begin{array}{c}
      r_1  \\
      r_2  \\
\end{array}\right)
=\left(\begin{array}{cc}
      -k\cos^2\theta_1 & -\Omega_I \cos\theta_1\cos\theta_2 \\
      \Omega_I \cos\theta_1\cos\theta_2 & -k\cos^2\theta_2 \\
\end{array}\right)
\left(\begin{array}{c}
      r_1  \\
      r_2  \\
\end{array}\right)\,.
\end{equation}
This looks familiar since we have almost the same equation as for
three-level system, cf. Eq.(\ref{eq:r1r2}).
We first observe that $\Omega_I$ should always take on the
maximal amplitude $A$: if $\Omega_I< A$, we can always increase it
to $A$ while lowering $\cos\theta_1$ and $\cos\theta_2$ such that the
rotation speed $\Omega_I \cos\theta_1\cos\theta_2$ remains the same,
but the effect of decay on $r_1$, $r_2$ is decreased.
 The problem therefore reduces to
\begin{equation}\label{eq:4level}
\frac{d}{dt}\left(\begin{array}{c}
      r_1  \\
      r_2  \\
\end{array}\right)
=\left(\begin{array}{cc}
      -ku_1^2 & - Au_1u_2 \\
         Au_1u_2 & -ku_2^2 \\
\end{array}\right)
\left(\begin{array}{c}
      r_1  \\
      r_2  \\
\end{array}\right)\,,
\end{equation}
where $u_1=\cos\theta_1$ and $u_2=\cos\theta_2$. The same dynamics
arise in Nuclear Magnetic Resonance and analytical solutions to this
control problem were obtained using the optimal control technique in
Ref.~\cite{Khaneja03}.
We describe the characteristics of the optimal pulse
sequences and refer to Ref.~\cite{Khaneja03} for more details.

Case I: If $T \leq {{\cot^{-1}(2 \xi)}\over{A}}$, where
$\xi=\frac{k}{A}$, then $u_1^*(t) = u_2^*(t) = 1$ throughout. That is,
we obtain a hard pump pulse at $t=0$, flipping the angle $\theta_1$ by
$\pi/2$, i.e. transferring all population from $x_1$ to $x_2$,
and a hard Stokes pulse at $t=T$, flipping $\theta_2$ by
$\pi/2$, transferring population from $x_3$ to $x_4$.
At intermediate times, the optimal Rabi frequencies for
pump and Stokes pulses are zero.  The efficiency of the population
transfer, $\eta_T$, is obtained by integrating Eq.~\eqref{eq:4level}
with the optimal controls $u_1^*(t) = u_2^*(t) = 1$,
\begin{equation}
  \eta_T = \exp[-kT] \sin(A\,T)\,.
\end{equation}
This is smaller than unity for all $T>0$.

Case II: For larger control times, $T > {{\cot^{-1}(2 \xi)}\over{A
  }}$, it is not optimal to put all population immediately into the
decaying level $x_2$. The optimal trajectory then has three distinct
phases.
\begin{figure}[tb]
  \centering
  \includegraphics[width=0.95\linewidth]{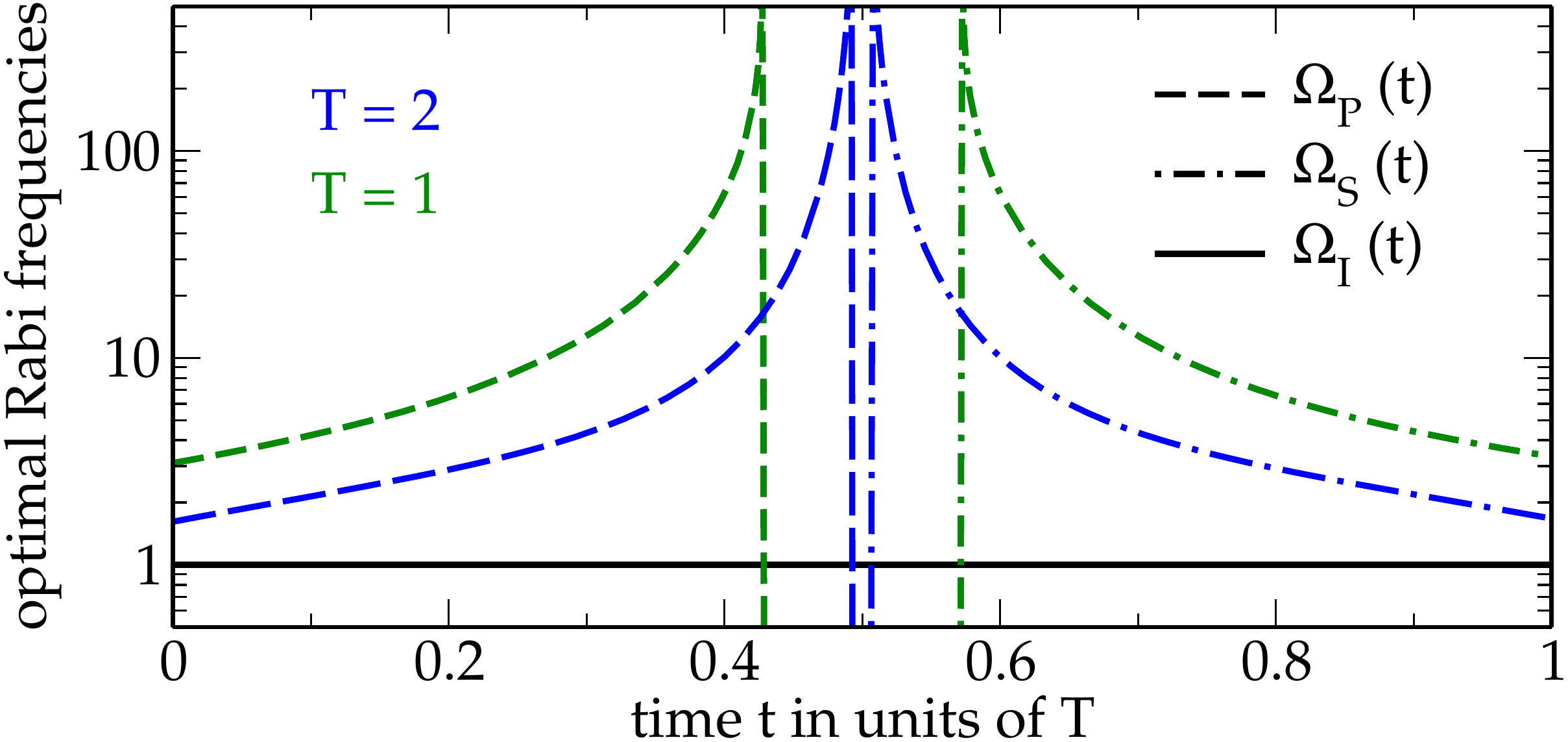}
  \caption{(Color online) Optimal half Rabi frequencies $\Omega_P(t)$,
    $\Omega_I(t)$, and
    $\Omega_S(t)$  (Case 2) for finite control times $T$
    with relaxation rate $k=1$  and the amplitude bound on
    $\Omega_I$ set to one, $A=1$. For very short $T$ (Case 1), the
    optimal Rabi frequencies $\Omega_S(t)$ and $\Omega_P(t)$
    correspond to instantaneous pulses of infinite power at $t=0$ and
    $t=T$; for $T \to\infty$, the hold period with zero
    $\Omega_S(t)$, $\Omega_P(t)$ in the middle of the
    time interval disappears.
    }
  \label{fig:fields-4level}
\end{figure}
For $0 \leq t \leq \tau$, where $\tau$ is a function of $T$,
$u_2^*(t) = 1$ and $u_1^*(t)$ is increased gradually from a value
$u_1^*(0) < 1$ to $u_1^*(\tau) = 1$.
This corresponds to $\Omega_p(t)$ rising from its initial value
$\Omega_p(0)$  to infinity while
$\Omega_s(t)$ remains zero throughout the first phase,
cf. Fig.~\ref{fig:fields-4level}.
In the second phase, for time $\tau \leq t \leq T-\tau$,
the optimal controls are $ u_1^*(t) = u_2^*(t) = 1$.
This corresponds to both  $\Omega_p(t)$ and $\Omega_s(t)$ being
zero.
Finally, for $t \geq T- \tau$, $u_1^*(t) = 1$ and
$u_2^*(t)$ is decreased from $u_2^*(T-\tau)=1$ to $u_2^*(T) = u_1(0)$.
$\Omega_s(t)$ is thus decreased from infinity to its final value
$\Omega_s(T)$ which
is equal to $\Omega_p(0)$, while $\Omega_p(t)$ remains zero.
The parameter $\tau$ determining the switching times
is calculated from the following equation,
\begin{equation}\label{eq:tau}
T = 2 \tau + {{ \gamma_2 - \gamma_1}\over{A}}\,,
\end{equation}
where
\[
\gamma_1 = \cot ^{-1}\left(\frac{1 - \kappa(\tau)}{2
\xi \kappa(\tau)}\right)\,, \ \
\gamma_2 = \tan ^{-1}\left(\frac{1
- \kappa(\tau)}{2\xi}\right)\,,
\]
and
\begin{eqnarray*}
\kappa(\tau) &=& 1 + 2 \xi^2 -\\
&&2 \xi \sqrt{1 + \xi^2}
\coth\left(A \sqrt{1 + \xi^2}\ \tau
+  2 \sinh^{-1}\xi\right)\,.
\end{eqnarray*}
In the limit of infinite time, $\tau = \frac{T}{2}$; in this case the 'hold'
phase in the middle, where only $\Omega_I(t)$ is non-zero, disappears
(cf. the blue and green curves in Fig.~\ref{fig:fields-4level}).
The optimal solution thus corresponds to the intuitive pulse sequence
of pump first, then Stokes, not a STIRAP-like solution characterized
by a counter-intuitive pulse sequence.

The efficiency of the population transfer, $\eta_T$, in case 2 is
expressed for finite time $T$ in terms of the angles as
\begin{equation}\label{eq:opt.u.3} \eta_T = \frac{\exp(\xi(\gamma_1 -
    \gamma_2))(1 - \xi \sin 2 \theta_2)}{\sin (\gamma_1 + \gamma_2)}\,.
\end{equation}
In the limit that $T$ goes to infinity,
$\gamma_1 = \gamma_2 = \tan^{-1}\left(\sqrt{1 + \xi^2}
  -\xi\right)$, and $\eta_T$ approaches $\eta$, the
maximum transfer efficiency given by
\begin{equation}\label{eq:efficiency}
\eta = \sqrt{1 + \xi^2} - \xi\,.
\end{equation}
We see that the transfer efficiency can reach unity  only for $\xi=0$,
i.e., $\frac{k}{A}=0$. This is the case of infinite power, i.e., the
decay is much smaller than the maximum Rabi frequency coupling the
intermediate states. When only limited power is
available, the transfer efficiency is always less than unity.

Even for the case of a multiphoton resonance, no analytical optimal
solutions are known for non-zero detunings.
As shown in Refs.~\cite{VitanovEPJD98,NakajimaPRA99},
adiabatically eliminating one of the two decaying intermediate levels
allows one to recover a three-level
system and thus the STIRAP solution. However, perfect adiabatic
elimination in the presence of decay requires infinite power of the
field coupling the intermediate levels.
We believe that our conclusion for a finite power
transfer efficiency of less than unity holds also for
non-zero detunings.

To support this claim, consider the four-level system with non-zero detunings
$\Delta_2$, $\Delta_3$,
\begin{equation}\label{eq:4levelwithdetuning}
\frac{d}{dt}\left(\begin{array}{c}
      x_1  \\
      x_2  \\
      x_3  \\
      x_4
\end{array}\right)
=-i\left(\begin{array}{cccc}
      0 & \Omega_p & 0 & 0\\
      \Omega_p & \Delta_2-ik & \Omega_i & 0 \\
      0 & \Omega_i & \Delta_3-ik & \Omega_s\\
      0 & 0 & \Omega_s & 0
\end{array}\right)
\left(\begin{array}{c}
      x_1  \\
      x_2  \\
      x_3  \\
      x_4
\end{array}\right)\,.
\end{equation}
Assume complete population transfer is possible in this
system. Clearly, for this to hold the population of the intermediate
states must remain zero during the process. If this is the case, the
values of the detunings would not affect the transfer and we can
therefore replace the detunings by zero. However, in this case the
system reduces to Eq.~\eqref{eq:4levelwithdetuning},
which we showed to be uncontrollable with
finite power, leading to a contradiction with our assumption of
controllability. Thus we claim that even with non-zero detunings,
complete population transfer is not possible with finite pulse power
if there are multiple decaying intermediate states. Since detuning
relative to the intermediate states is
known not to affect STIRAP efficiency \cite{BergmannRMP98}, this
result is fully consistent with our prior expectations.

\section{Generalization to $N$-level systems}\label{sec:N}
In the previous section we showed that with two consecutive
intermediate decaying states, the transfer efficiency with finite
power is always less than one.
If more than two intermediate decaying states are present, the transfer
efficiency  gets even
worse. For example, suppose we have a chain of five states with three
intermediate decaying states,
\[
\frac{d}{dt}\left(\begin{array}{c}
      x'_1  \\
      x'_2  \\
      x'_3  \\
      x'_4  \\
      x'_5
\end{array}\right)
=-i\left(\begin{array}{ccccc}
      0 & \Omega_p & 0 & 0 & 0\\
      \Omega_p & -ik & \Omega_1 & 0 &0 \\
      0 & \Omega_1 & -ik & \Omega_2 &0\\
      0 & 0 & \Omega_2 & -ik &\Omega_s\\
      0 & 0 & 0 & \Omega_s & 0
\end{array}\right)
\left(\begin{array}{c}
      x'_1  \\
      x'_2  \\
      x'_3  \\
      x'_4   \\
      x'_5
\end{array}\right)\,.
\]
With a change of variables, letting $x_1=x_1'$, $x_2=ix'_2$,
$x_3=-x_3'$, $x_4=-ix_4'$ and $x_5=x'_5$, the dynamics become
\[
  \frac{d}{dt}\left(\begin{array}{c}
      x_1  \\
      x_2  \\
      x_3  \\
      x_4   \\
      x_5
    \end{array}\right)
  =\left(\begin{array}{ccccc}
      0 & -\Omega_p & 0 & 0 &0\\
      \Omega_p & -k & -\Omega_1 & 0 &0\\
      0 & \Omega_1 & -k & -\Omega_2 &0 \\
      0 & 0 & \Omega_2 & -k & -\Omega_s \\
      0 & 0 & 0 & \Omega_s & 0
    \end{array}\right)
  \left(\begin{array}{c}
      x_1  \\
      x_2  \\
      x_3  \\
      x_4  \\
      x_5
\end{array}\right)\,.
\]
Introducing new variables $y_i$, the fifth state and fourth state can
be combined,
\[
y_1=x_1\,,\,
y_2=x_2\,,\,
y_3=x_3\,,\,
y_4=\sqrt{x_4^2+x_5^2}\,.
\]
This reduces the system to an effective four-state system,
\begin{equation}\label{eq:dyna5}
  \frac{d}{dt}\left(\begin{array}{c}
      y_1  \\
      y_2  \\
      y_3  \\
      y_4
    \end{array}\right)
  =\left(\begin{array}{cccc}
      0 & -\Omega_p & 0 & 0\\
      \Omega_p & -k & -\Omega_1 & 0 \\
      0 & \Omega_1 & -k & -\Omega_2\cos\theta \\
      0 & 0 & \Omega_2\cos\theta & -k\cos^2\theta
    \end{array}\right)
  \left(\begin{array}{c}
      y_1  \\
      y_2  \\
      y_3  \\
      y_4
\end{array}\right)\,,
\end{equation}
where $\tan\theta=\frac{x_5}{x_4}$.
The transfer efficiency for the dynamics of $y_1,\ldots,y_4$,
represents an upper bound to the transfer
efficiency for the dynamics of $x_1,\ldots,x_5$, since $y_4\geq x_5$:
If a control scheme transfers an amount of population, $\eta$,
from $x_1$ to $x_5$ then this scheme also transfers at least an amount
of population $\eta$ from $y_1$ to $y_4$.
Next we show that the efficiency of transfer from $y_1$ to $y_4$
according to the dynamics of Eq.~\eqref{eq:dyna5}
is upper bounded by the transfer efficiency for the
dynamics of Eq.~(\ref{eq:dyna4}) of the four-level chain. To make the
comparison transparent, we consider the following dynamics,
\begin{equation}\label{eq:inter}
  \frac{d}{dt}\left(\begin{array}{c}
      y'_1  \\
      y'_2  \\
      y'_3  \\
      y'_4
    \end{array}\right)
  =\left(\begin{array}{cccc}
      0 & -\Omega_p & 0 & 0\\
      \Omega_p & -k & -\Omega_1 & 0 \\
      0 & \Omega_1 & -k & -\Omega_2\cos\theta \\
      0 & 0 & \Omega_2\cos\theta & 0
    \end{array}\right)
  \left(\begin{array}{c}
      y'_1  \\
      y'_2  \\
      y'_3  \\
      y'_4
\end{array}\right)\,.
\end{equation}
Clearly the efficiency of transfer from $y_1$ to $y_4$ is upper bounded
by the efficiency of transfer from $y'_1$ to $y'_4$ since the only
difference between these two dynamics is that $y_4$ is subject to
decay while $y'_4$ is not.
The efficiency of transfer from $y'_1$ to $y'_4$ in turn is
upper bounded by the transfer efficiency for the four-level chain,
Eq.~(\ref{eq:dyna4}),
since the controls in Eq.~(\ref{eq:inter}) are more restricted than
the controls in  Eq.~(\ref{eq:dyna4}). This can be seen as follows: If
a control scheme for the dynamics, Eq.~(\ref{eq:inter}),
reaches an efficiency $\eta$, then simply setting
$\Omega_s=\Omega_2\cos\theta$ in Eq.~(\ref{eq:dyna4}),
the same transfer efficiency $\eta$ is obtained for the four-level
chain. The inverse step of setting
$\Omega_2=\frac{\Omega_s}{\cos\theta}$ is, however, not always possible,
since this may lead to infinite $\Omega_2$.

From this line of argument, we see that the transfer efficiency of the
five-state chain with three intermediate decaying states is upper
bounded by the transfer efficiency of the four-state chain with two
intermediate decaying states.
Analogously, we can show that the transfer efficiency of the $N$-state
chain $(N\ge 4)$ with $N-2$ intermediate decaying states
is upper bounded by the transfer efficiency of the $N$-state
chain with  $N-3$ intermediate decaying states. That is for chains
where all intermediate states are subject to decay, the transfer
efficiency is monotonically decreasing with increasing length of the
chain. In particular, we have the interesting result that
for any chain with two or more consecutive intermediate decaying states,
the efficiency will be less than unity.

Our analytical results allow us to draw a general inference about
previous work on population transfer in $N$-level systems.
Extensions of STIRAP from the three-level system to multi-level chains
have been investigated since
the early days of STIRAP (see Ref.~\cite{BergmannRMP98} and
references therein).
Although these mechanisms are designed to keep the population in the
intermediate levels as small as possible, close inspection reveals
that none of these succeed in completely avoiding intermediate
population for finite power pulses, even for $T \to\infty$. In
particular, we note that none of these schemes can keep the population
for two consecutive intermediate states at zero. We illustrate this
with three of the generalized STIRAP schemes.
In Ref.~\cite{ShorePRA91}, a generalization of STIRAP for $N$-level
chains was proposed for $N$ odd. In this scheme, the population in all
the intermediate even levels can be kept at zero but placing
population in the intermediate odd levels cannot be avoided. Clearly
it is impossible to keep two neighbouring states empty. In
Ref.~\cite{Tannor97}, a STIRAP-like solution for $N$-level chains was
found numerically, and termed straddling
STIRAP. It consists in choosing the Rabi frequencies coupling the
intermediate states to be at least one order of magnitude larger than
$\Omega_s$ and $\Omega_p$ and overlapping in time with both $\Omega_s$
and $\Omega_p$. Inspection of the solution reveals that all
intermediate levels acquire some population, and therefore it is
impossible to keep two consecutive levels empty without using infinite
power. The straddling STIRAP was analyzed further both
analytically and numerically \cite{VitanovEPJD98,NakajimaPRA99}.
It was clarified that for $N$ even, a non-zero detuning of the lasers
coupling the intermediate states is required to obtain a STIRAP-like
solution while for zero detunings, an intuitive (Rabi) pulse sequence
is found.
Moreover, it was shown that in the dressed state picture, a very
strong coupling between the intermediate states moves the
intermediate states out of resonance such that they are decoupled and
effectively a two-level system ($N$ even with zero detuning)
or a three-level system ($N$ odd or $N$ even with non-zero detuning)
are recovered.
It can be seen that whether $N$ is odd or even, avoiding population in
two consecutive levels requires infinite power.
Taking dissipation explicitly into account, it is clear that if two
consecutive intermediate levels are subject to decay, unit transfer
efficiency is impossible at finite power.
Our analytical results including dissipation
are consistent with this analysis.

\section{Controllability in the relaxation-free subspace}\label{sec:control}

In this section, we will generalize the results of the previous
sections to state-to-state controllability on
relaxation-free subspaces $S$. 
Based on the results of the previous sections, we will
characterize the relaxation-free subspaces that are finite-power controllable on
the pure state space. Our argument is based on the assumption of
selective control, i.e. the assumption that $\Omega_s$, $\Omega_p$ and
any other coupling, can  be tuned independently. This corresponds to
the bare, field-free Hamiltonian having no degenerate levels or
frequencies.  With selective control,
controllability on the relaxation-free subspace
becomes equivalent to connectivity~\cite{PolackPRA09}.

\begin{figure}[tb]
  \begin{center}
   \includegraphics[width=0.95\linewidth]{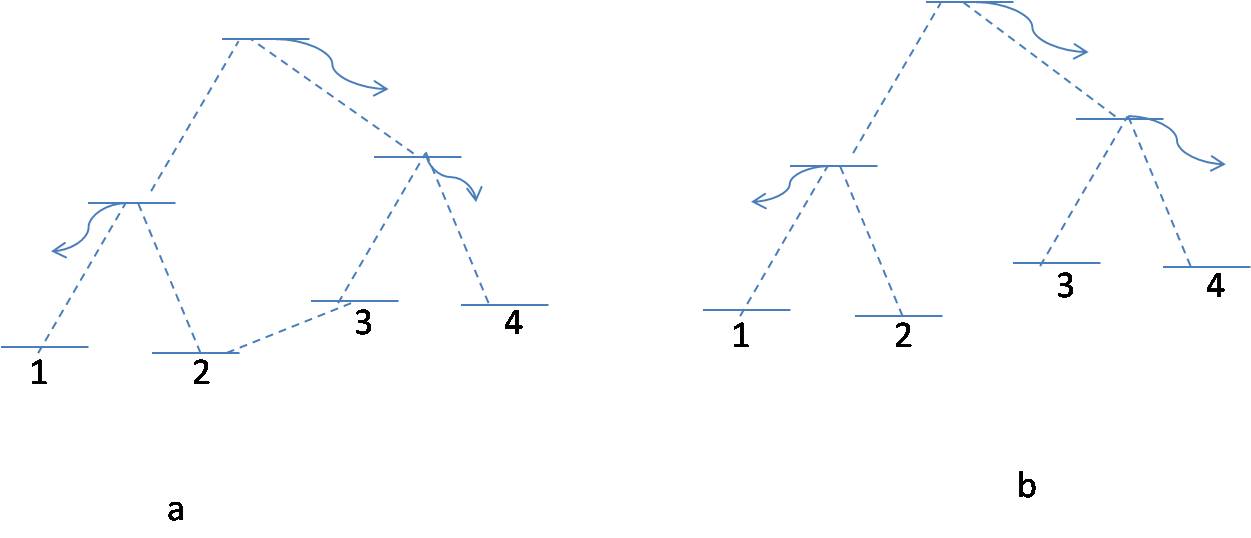}
  \end{center}
  \caption{(Color online)
    Examples for (a) a system that is controllable in the subspace
    $\{|1\rangle,|2\rangle,|3\rangle,|4\rangle\}$ since any pair of
    eigenstates can be connected by a path that never visits two
    consecutive states that suffer relaxation; (b) a system that is
    not  controllable in the subspace
    $\{|1\rangle,|2\rangle,|3\rangle,|4\rangle\}$ since
    one has to pass through three
    consecutive states that suffer relaxation
    to connect the states $|1\rangle$ and $|4\rangle$.
  }
  \label{fig:2}
\end{figure}

Our central result is that
 a relaxation-free subspace is controllable on the pure state
space \textit{if and only if any two
eigenstates in the subspace can be connected by a path that never
visits two consecutive states that both suffer relaxation}. This is
illustrated in Fig.~\ref{fig:2}
with part (a) displaying an example where
any pair of eigenstates in the relaxation-free subspace
$S=\{|1\rangle,|2\rangle,|3\rangle,|4\rangle\}$ can be connected by a
path that never visits two consecutive states suffering
relaxation, i.e. this system is state-to-state controllable in the space
$S=\{|1\rangle,|2\rangle,|3\rangle,|4\rangle\}$. The system shown in
Fig.~\ref{fig:2}(b) is not controllable in  the relaxation-free subspace
$S=\{|1\rangle,|2\rangle,|3\rangle,|4\rangle\}$
since one has to pass through
three consecutive states that suffer relaxation in order to connect
the states $|1\rangle$ and $|4\rangle$.

Clearly the condition is necessary. As from the previous section, we know
that if any
path has two intermediate states outside of the relaxation-free space,
complete population transfer is not possible. Hence the system is not
controllable.
The condition is also sufficient.
 If there are never two states in a row
that suffer relaxation, the control found in Section~\ref{sec:three}
allows us to traverse one
intermediate relaxing state without losses. Concatenating such processes
gives the result.

\begin{figure}[tb]
  \begin{center}
   \includegraphics[width=0.6\linewidth]{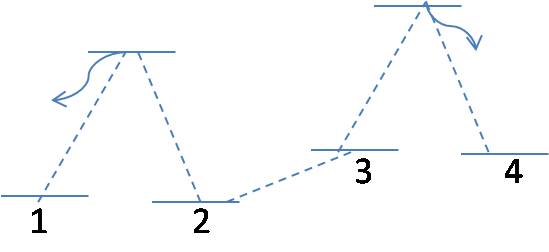}
  \end{center}
  \caption{(Color online)
    The eigenstates $|1\rangle$ and $|4\rangle$ are connected
    by concatenating paths of type (I) and type (II).
  }
  \label{fig:concat}
\end{figure}

We can connect the controllability condition with the coupling
topology of the system: The condition is  fulfilled if
and only if the eigenstates in the relaxation-free subspace
are (I) directly connected via paths in the subspace;
(II) connected by one intermediate state which suffers relaxation;
(III) connected by concatenations of paths of type I and type II as
sketched in Fig.~\ref{fig:concat}.
Note that this coupling topology includes degenerate levels in the
relaxation-free and the relaxing subspaces provided that a generalized
Morris-Shore transformation can be employed to replace the coupled
multi-level system by a set of  two- and three-level systems and dark
states (single levels)~\cite{RangelovPRA06}.

We first show that with this coupling
topology we can achieve coherent population transfer between any two
eigenstates. Controllability is then achieved by applying sequences of
these operations.
First, if two eigenstates are connected via (I)
we can obviously achieve coherent population transfer between them.
It remains to be shown that coherent transformation is
possible for two eigenstates that are connected by one intermediate
state suffering relaxation.
This is achieved by combining our results of
Section~\ref{sec:three} with the fractional STIRAP developed by
Vitanov et al.~\cite{Vitanov99} to generate arbitrary
coherent superpositions of $|1\rangle$ and $|3\rangle$ from the initial
state $|1\rangle$.  For two eigenstates $|1\rangle$
and $|3\rangle$ that are connected by an intermediate state $|2\rangle$
suffering relaxation, a coherent transformation
from state $|1\rangle$ to
$\cos\beta|1\rangle-e^{i\phi}\sin\beta|3\rangle$ is
implemented by (i) adding a phase $\phi$ to the
Stokes pulse such that the equations of motion read
  \begin{equation}
  \frac{d}{dt}\left(\begin{array}{c}
      x'_1  \\
      x'_2  \\
      x'_3
    \end{array}\right)
  =-i\left(\begin{array}{ccc}
      0 & \Omega_p & 0 \\
      \Omega_p & -ik & \Omega_se^{-i\phi} \\
      0 & \Omega_se^{i\phi} & 0
    \end{array}\right)
  \left(\begin{array}{c}
      x'_1  \\
      x'_2  \\
      x'_3
    \end{array}\right)\,,
\end{equation}
and (ii) varying $\frac{\Omega_p}{\Omega_s}$ adiabatically from $0$
to $\tan\beta$.
A general coherent transformation can be implemented by first applying
the time-reversed version of this procedure in order to
transfer the initial state to $|1\rangle$ and then to transfer
$|1\rangle$ to the desired final state with the control scheme of
Section~\ref{sec:three}.
We have thus shown controllability for
eigenstates connected by (II). Obviously the results for (I) and
(II) can be combined which yields controllability for (III) completing
the proof of sufficiency.

So far we have considered only zero detunings. However, the result of
controllability on the relaxation-free subspace also holds for
non-zero detunings that can be represented by complex
values for the decays.
To see this, consider first a three-level
system. It is well-known that the values of the detuning and the
relaxation rate of the intermediate state do not affect the STIRAP
efficiency~\cite{BergmannRMP98}.
This argument carries over to general N-level systems
with non-zero detunings of the decaying states. If the system is
controllable it can be viewed as a concatenation of two or more
three-level STIRAP systems (and possibly type (I)
paths), and then the detuning is irrelevant. If the system is
uncontrollable, the detuning cannot make it controllable, based on the
argument presented at the end of Section~\ref{sec:four}.

\section{Discussion and conclusions}\label{sec:concl}
We have considered the optimal control problem of transferring
population in a
quantum system between states in a subspace free of dissipation,
where the transfer has to proceed via states that are subject
to decay. We treated only the case of resonant controls with fixed
carrier frequency and phases, controlling only the amplitudes of our
pulses as a function of time.
Such situations occur frequently in atomic and molecular
physics applications. For example, transfer between different levels
in the electronic ground manifold can proceed via Raman transitions. In
quantum information applications, stable qubit states are often
connected via auxiliary states that are subject to decay. In
particular, this may be the case for logical qubits encoded in a
decoherence-free subspace.

We have obtained analytical solutions to this optimal control problem
by solving the Hamilton-Jacobi-Bellman equation for the optimal return
function.
For a single intermediate decaying state, we have recovered
the Stimulated Raman Adiabatic  Passage process \cite{BergmannRMP98}
as the globally optimal solution
in the limit of infinite time. Perfect state transfer is achieved only
in this limit. This is in accordance with experimental realizations
where at best 99.5\% state transfer were achieved \cite{BergmannPC}.

In Ref. \cite{Boscain02}, the STIRAP solution in a three-level system
was previously obtained using geometric control methods. There,
the optimal return function was specifically designed to avoid the hard
pulses obtained by us (which cause the sudden population transfer from
level 1 to 2 at $t=0$). Note that generalizing the results
of Ref. \cite{Boscain02} from the
three-level system to $N$-level systems is hampered by the
system's state being represented in terms of six real variables,
compared to two variables, $r_1$ and $r_2$, in
our case.

In contrast to the analytical solutions
presented here and in Ref. \cite{Boscain02},
Refs. \cite{BandJCP94,WangJCP96,TannorPRA99,KisJMO02} employ numerical
optimization procedures based on the calculus of variations.
Our current work may help to clarify the disagreement in the literature
on whether STIRAP is obtained as a solution to an optimal control
problem \cite{BandJCP94,Tannor97,TannorPRA99}: the assertion of
Ref. \cite{BandJCP94} that adiabatic passage population transfer
cannot be obtained as the solution to an optimal control problem
implicitly assumes finite pulse fluence and finite control
time. Yet adiabaticity, strictly speaking, does not comply with these
assumptions.

We have also presented analytical solutions for the case of population
transfer that proceeds via two consecutive intermediate decaying
states. In particular, we have shown that the optimal
control does not yield perfect state transfer even in the limit of
infinite time, unless the pulse coupling the intermediate levels has
infinite power. This gives an analytical framework for understanding
an earlier control solution termed straddling STIRAP that was
obtained numerically \cite{Tannor97} and that is essentially based on
adiabatically eliminating the intermediate levels
\cite{VitanovEPJD98,NakajimaPRA99}. Taking dissipation explicitly into
account, we have clarified that the adiabatic
elimination of the decaying levels is possible only in the limit of
infinite power.

Finally, we have generalized these
results to characterize the topologies of paths that can be achieved
in $N$-level systems by finite-power controls and in the presence of
dissipation. Population transfer with
unit efficiency is only possible if each decaying state is connected
to two non-decaying states. Complete population transfer is then
achieved in the adiabatic limit, i.e., in a sequence
of STIRAP processes. Finite-power state-to-state controllability on
the relaxation-free subspace is thus equivalent
to connectivity~\cite{PolackPRA09}, augmented by the condition that
only one out of two consecutive levels may be subject to dissipation.

\begin{acknowledgments}
  We enjoyed the hospitality of the KITP in the framework of the
  Quantum Control of Light and Matter program (KITP
  preprint No. NSF-KITP-10-016, this  research was
  supported in part by the National Science Foundation  under Grant
  No. PHY05-51164).
  CPK is grateful to the Deutsche
  Forschungsgemeinschaft for financial support (Grant No. KO
  2301/2). DJT acknowledges
  financial support from the Minerva Foundation with funding from the
  Federal Ministry for Education and Research.
\end{acknowledgments}

\appendix
\section{Optimal control for the three-level system}\label{sec:optimal}
In this Appendix, we derive the solution of the optimal control problem for
the three level system. As introduced in Section~\ref{sec:three}, we
are going to solve the
Hamilton Jacobi Bellman equation, Eq.~(\ref{eq:HJB}) with the
classical Hamiltonian of the control problem defined in Eq.~(\ref{eq:Ham}).
Introducing adjoint variables $\lambda_1$, $\lambda_2$,
\begin{equation}
  \label{eq:lambdas}
\lambda_1=\frac{\partial V}{\partial r_1} \,,\,
\lambda_2=\frac{\partial V}{\partial r_2}\,,
\end{equation}
where $V$ denotes the optimal return function,
the Hamiltonian of the optimal control problem can be expressed as
\begin{widetext}
\bea
\label{eq:newH}
\aligned
H(u)&=-k\lambda_1r_1u^2-A\lambda_1r_2u+A\lambda_2r_1u
=-k\lambda_1r_1\left[u^2+\frac{A}{k}\left(\frac{r_2}{r_1}-
  \frac{\lambda_2}{\lambda_1}\right)u\right]\\
&=-k\lambda_1r_1\left\{\left[u-\frac{A}{2k}\left(\frac{\lambda_2}{\lambda_1}-\frac{r_2}{r_1}\right)\right]^2-\left[\frac{A}{2k}\left(\frac{\lambda_2}{\lambda_1}-\frac{r_2}{r_1}\right)\right]^2\right\}\,.
\endaligned
\eea
\end{widetext}
We rewrite the ratios appearing in Eq.~(\ref{eq:newH}) in terms of variables $a$
and $b$,
\begin{equation}
  \label{eq:def_ab}
  a=\frac{\lambda_2}{\lambda_1} \,,\, b=\frac{r_2}{r_1}  \,.
\end{equation}
The optimal
return function is a non-decreasing function of
$r_1$, $r_2$ [starting from a larger $r_1(0)$ or $r_2(0)$, one can achieve
a larger $r_2(T)$]. Due to Eq.~(\ref{eq:lambdas}) we therefore find
$\lambda_1 \ge 0$, $\lambda_2\ge0$ and hence $a\geq 0$.
Since $\lambda_1\geq0$ and $r_1\geq0$, maximizing $H$ with respect to
$u$ is equivalent to minimizing the function
$$
f(u)=\left[u-\frac{A}{2k}(a-b)\right]^2\,.
$$
If $a-b<0$, then the solution is the trivial one, $u^*=0$. We
therefore conclude that obtaining a non-zero control requires
$a-b\geq 0$. Later we will show explicitly that $a-b$ is a
non-decreasing function of time. Since $b(0)=0$ and $a(0)\geq 0$, we have
$a(t)-b(t)\geq 0$ for all times $t$. We further distinguish two cases.

Case I: If $\frac{A}{2k}(a-b)\geq 1$, then the minimum of $f(u)$ is
achieved at the maximum value that $u=\cos\theta$ can take, $u^*=1$.

Case II: If $0\leq \frac{A}{2k}(a-b)<1$, then the minimum of $f(u)$ is
achieved at $u^*=\frac{A}{2k}(a-b)$.

It is a standard result that, along the optimal trajectory
$(r_1(t),r_2(t))$, the adjoint variables $(\lambda_1(t),\lambda_2(t))$
satisfy the equations
\begin{eqnarray*}
  \frac{d\lambda_1}{dt}&=&-\frac{\partial H}{\partial r_1}\,,\\
  \frac{d\lambda_2}{dt}&=&-\frac{\partial H}{\partial r_2}\,,
\end{eqnarray*}
i.e.,
\begin{equation}\label{eq:adjoint}
\frac{d}{dt}\left(\begin{array}{c}
      \lambda_1  \\
      \lambda_2  \\
\end{array}\right)
=\left(\begin{array}{cc}
      ku^2 & -A u \\
      A u & 0 \\
\end{array}\right)
\left(\begin{array}{c}
      \lambda_1  \\
      \lambda_2  \\
\end{array}\right)
\end{equation}
with the terminal condition $\lambda_1(T)=0$, $\lambda_2(T)=1$.

With Eqs.~(\ref{eq:adjoint}) and~(\ref{eq:r}), we can derive
the dynamics for $a$ and $b$ along the optimal trajectory,
\begin{subequations}\label{eq:ab}
\bea
\frac{d}{dt}a&=Aua^2-ku^2a+Au \,,\\
\frac{d}{dt}b&=Aub^2+ku^2b+Au \,.
\eea
\end{subequations}
Therefore
\be
\label{eq:a-b}
\frac{d}{dt}(a-b)=(a+b)u[A(a-b)-ku] \,.
\ee

For the optimal trajectory starting at $(r_1,r_2)=(1,0)$, $b(0)=0$.
Depending on $a(0)$ we have the following cases.

\begin{figure}[tb]
  \centering
  \includegraphics[width=0.95\linewidth]{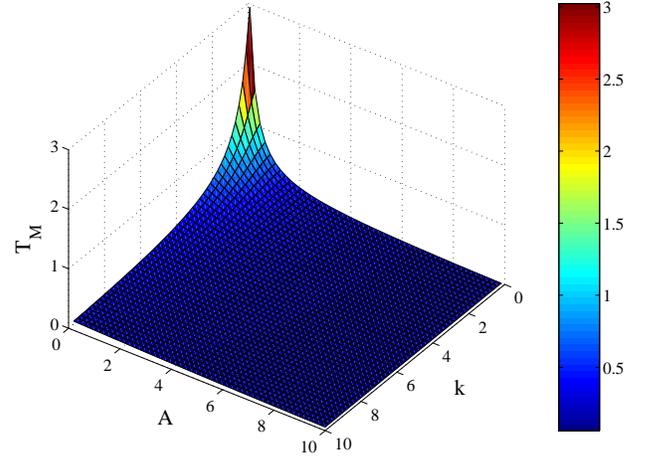}
  \caption{(Color online)
    The critical time $T_M$ as a function of decay rate $k$
    and amplitude bound $A$. For control times $T$ shorter than $T_M$ the
    optimal control is simply set to $u^*=1$ for all $t \in [0,T]$,
    for control times larger than
    $T_M$, Eq.~(\ref{eq:uoptimal}) applies.}
  \label{fig:TM}
\end{figure}

Case A: If $a(0)\geq \frac{2k}{A}$, we start in Case I,
$u^*=1$. From equation~(\ref{eq:a-b}), it follows that
$$
\frac{d}{dt}(a-b)=(a+b)[A(a-b)-k]\geq k(a+b)\geq 0\,,
$$
i.e., $a-b$ is non-decreasing. Therefore we will remain in Case I for
the whole time interval. Substituting $u^*=1$ into the dynamical equation
for $(\lambda_1,\lambda_2)$, Eq.~(\ref{eq:adjoint}), and running it
backwards, we obtain
\begin{equation}\label{eq:adjointback}
\frac{d}{dt}\left(\begin{array}{c}
      \lambda_1(T-t)  \\
      \lambda_2(T-t)  \\
\end{array}\right)
=\left(\begin{array}{cc}
      -k & A  \\
      -A  & 0 \\
\end{array}\right)
\left(\begin{array}{c}
      \lambda_1(T-t)  \\
      \lambda_2(T-t)  \\
\end{array}\right)
\end{equation}
with the initial condition $\lambda_1(T)=0$,
$\lambda_2(T)=1$. Integrating this equation yields
\begin{subequations}\label{eq:lambdas_t}
  \begin{widetext}
\bea
\label{eq:back}
\left(\begin{array}{c}
    \lambda_1(t) \\
    \lambda_2(t)
  \end{array}\right)=
\left(\begin{array}{c}
    \frac{1}{\omega}
    \left[-Ae^{-\frac{1}{2}(k+\omega)(T-t)}+Ae^{-\frac{1}{2}(k-\omega)(T-t)}\right]\\
    \frac{1}{2\omega}\left[
      -e^{-\frac{1}{2}(k+\omega)(T-t)}(k-\omega)+
      e^{-\frac{1}{2}(k-\omega)(T-t)}(k+\omega)\right]
  \end{array}\right)\,,
\eea
\end{widetext}
where
\begin{displaymath}
\omega = \left\{ \begin{array}{ll}
\sqrt{k^2-4A^2} & \textrm{if $k^2>4A^2$}\\
i\sqrt{4A^2-k^2} & \textrm{if $k^2<4A^2$}\\
\end{array} \right.
\end{displaymath}
and
\bea
\label{eq:k=2A}
\left(\begin{array}{c}
      \lambda_1(t) \\
      \lambda_2(t)
\end{array}\right)=
\left(\begin{array}{c}
   Ae^{-A(T-t)}(T-t)\\
   e^{-A(T-t)}[A(T-t)+1]
\end{array}\right)
\eea
\end{subequations}
for $k^2=4A^2$.
Solving the equation
$a(0)=\frac{\lambda_2(0)}{\lambda_1(0)}=\frac{2k}{A}$ for $T$, we
obtain a critical time, $T_M$. When $T\leq T_M$, the optimal control is
to set $u^*=1$ for the whole time $[0,T]$. An analytical
expression can be derived for $T_M$. For example, when $k^2>4A^2$,
\be
\label{eq:TM}
T_M=\frac{\log\left[\frac{-2A^2+5k^2+3k\sqrt{k^2-4A^2}}{2(A^2+2k^2)}\right]}
{\sqrt{k^2-4A^2}} \,.
\ee
The other cases can also be worked out.
The critical time $T_M$ as a function of the decay rate $k$ and the
amplitude bound $A$ is displayed in Fig.~\ref{fig:TM}.
$T_M$ takes large values corresponding to the trivial optimal solution
$u^*=1$ for all $t \in [0,T]$
only for small decay rates and small amplitude bounds. As $k$
and $A$ increase, $T_M$ quickly becomes  fairly small (for example, $T_M=0.06$
for $k=A=10$) and the optimal solution
is determined according to Eq.~(\ref{eq:uoptimal}).

Case B: If $a(0)< \frac{2k}{A}$,  we start in Case II,
$u^*=\frac{A}{2k}(a-b)$. From Eq.~(\ref{eq:a-b}), we get in
this case
\bea
\label{eq:u}
\frac{d}{dt}(a-b)&=&(a+b)\frac{A}{2k}(a-b)\left[A(a-b)-\frac{A}{2}(a-b)\right]
\nonumber \\
&=&\frac{A^2}{4k}(a+b)(a-b)^2\geq 0
\eea
If $a(0)-b(0)=0$, then it will remain zero for the whole time.
We will see that this only occurs as $T\rightarrow
\infty$. If $a(0)>0$, then $a(t)-b(t)$ is strictly increasing, and at some time
$\tau$, it will reach $\frac{2k}{A}$. We then switch to Case
I, setting $u^*=1$ afterwards. So in this case, the optimal control is
$u^*(t)=\frac{A}{2k}[a(t)-b(t)]$ for $t\leq \tau$ and $u^*(t)=1$ for
$\tau<t\leq T$.

We now show how to calculate the switching time $\tau$ of Case B.
Using Eqs.~(\ref{eq:ab}) and $u^*=\frac{A}{2k}(a-b)$, one can show
that within the time interval $[0,\tau]$,
\be
\label{eq:a-ba+b}
\frac{d}{dt}\left(\frac{a-b}{a+b}\right)
=-\frac{A^2}{k}\left(\frac{a-b}{a+b}\right)^2\,.
\ee
Together with the initial condition, $\frac{a(0)-b(0)}{a(0)+b(0)}=1$,
keeping in mind that  $b(0)=0$, this yields
\be
\label{eq:tau2}
\frac{a(\tau)-b(\tau)}{a(\tau)+b(\tau)}=\frac{1}{\frac{A^2}{k}\tau+1} \,.
\ee
Since at time $t=\tau$, $a(\tau)-b(\tau)=\frac{2k}{A}$, we obtain
\begin{subequations}\label{eq:ab_tau}
\bea
a(\tau)&=&A\tau+\frac{2k}{A} \label{eq:a_tau} \,,\\
b(\tau)&=&A\tau \,.
\eea
\end{subequations}
In the time interval $[\tau, T]$, $u^*=1$, and we can again run
Eq.~(\ref{eq:adjointback}) for $(\lambda_1,\lambda_2)$ backwards in
time from $T$ to $\tau$. This yields another expression of $a(\tau)$ from
Eqs.~(\ref{eq:back}) and~(\ref{eq:k=2A}). Setting it equal to
$A\tau+\frac{2k}{A}$, we obtain the switching time $\tau$.

Note that $T-\tau \le T_M$. From Eqs.~(\ref{eq:ab_tau}), the fact
that $b(t)$ is non-negative for all times $t$ and $a(t)-b(t)$ is an
increasing function, it follows that
\[
a(t) > \frac{2k}{A} \quad\mathrm{for}\quad t\in[\tau,T]\,.
\]
The assumption  $T-\tau > T_M$ then leads to a contradiction: If
$T-\tau > T_M$, then $a(T-T_M)=2k/A$ at time $t=T-T_M$.

Next we evaluate the value of the optimal control
$u^*(t)=\frac{A}{2k}[a(t)-b(t)]$ for $t\leq \tau$. We know that in the
interval $[0,\tau]$, $a-b$ satisfies the dynamical equation,
Eq.~(\ref{eq:u}), which can be rewritten
\bea
\frac{d}{dt}(a-b)=\frac{A^2}{4k}\frac{a+b}{a-b}(a-b)^3 \,.
\eea
From Eq.~(\ref{eq:tau2}), we get $\frac{a+b}{a-b}=\frac{A^2}{k}t+1$.
Substituting it into the above differential equation and solving it,
we obtain $a(t)-b(t)$ for $t\in[0, \tau]$,
\be
a(t)-b(t)=\frac{2k}{A}\frac{1}{\sqrt{A^2(\tau^2-t^2)+2k(\tau-t)+1}}\,.
\ee
So in Case B, the optimal control is obtained to be
\bea
\label{eq:control}
u^*(t) = \left\{ \begin{array}{ll}
\frac{1}{\sqrt{A^2(\tau^2-t^2)+2k(\tau-t)+1}} & \textrm{for $t\in[0,\tau]$}\\
1 & \textrm{for $t\in[\tau,T]$}\\
\end{array} \right. \,.
\eea
In summary, there is a critical time, $T_M$, which depends on the
relaxation rate $k$ and the amplitude bound $A$ and
determines whether the control is switched or not:
For $T\leq T_M$ the optimal control is just set to one, $u^*(t)=1$,
for all times $t\in[0,T]$, and for $T>T_M$,
\begin{displaymath}
u^*(t) = \left\{ \begin{array}{ll}
\frac{1}{\sqrt{A^2(\tau^2-t^2)+2k(\tau-t)+1}} & \textrm{for $t\in[0,\tau]$}\\
1 & \textrm{for $t\in[\tau,T]$}\\
\end{array} \right.\,.
\end{displaymath}
The optimal control $u^*(t)$ is thus determined by the system
parameters $A$ and $k$ and the switching time $\tau$ which is
obtained by matching the dynamics of
$a(t)$ at $t=\tau$, cf. Eqs.~(\ref{eq:a_tau}) and~(\ref{eq:lambdas_t}).

We derive the optimal Rabi frequency, $\Omega_p(t)$, from
$r_1(t)$, $r_2(t)$ and $u^*(t)$. From Eq.~(\ref{eq:dyna}), it follows
that
\be\label{eq:x_2}
\frac{d}{dt}x_2=\Omega_px_1-kx_2-\Omega_sx_3 \,.
\ee
Substituting in Eq.~(\ref{eq:x_2}) $x_2=r_1u$, $x_1=r_1\sqrt{1-u^2}$,
and $x_3=r_2$, we obtain
\be
\label{eq:omega}
\Omega_p=\frac{-ku^2r_1-Aur_2+r_1\dot{u}+kr_1u+A r_2}{r_1\sqrt{1-u^2}}\,.
\ee
It is difficult to obtain a closed form of $\Omega_p(t)$, but
Fig.~\ref{fig:fields} shows a few numerical examples.
The solution $u^*(t)=1$ occurs toward the end of the interval
$[0,T]$. For finite control times $T$, this solution for $u^*(t)$
corresponds to $\Omega_p(t)$ being infinity. For $T \to\infty$, a
rescaling of time leads to finite $\Omega_p(t)$ as explained in
Section~\ref{subsec:stirap}.

The interpretation of the optimal pulses is as follows:
For small control time $T$, the major limitation for the population
transfer is not due to relaxation, but the
limited available time. The optimal choice $u^*=1$ maximizes
the transfer speed,  but also
maximizes the decay of $r_1(t)$, respectively $x_2(t)$,
as can be seen from Eq.~(\ref{eq:r}). However,
for a small available control time $T$, the gain obtained by maximizing
the desired transfer at each moment is more important than
the relaxation losses. As the control time $T$ increases, the relaxation
degrades the performance more and more and the choice $u=1$ ceases to
be optimal. For finite time the optimal solution becomes a compromise
between maximizing the transfer speed and minimizing the decay. When
time goes to infinity, minimizing the relaxation loss becomes more
important than the transfer speed.

\begin{figure}[tb]
\begin{center}
\includegraphics[width=0.95\linewidth]{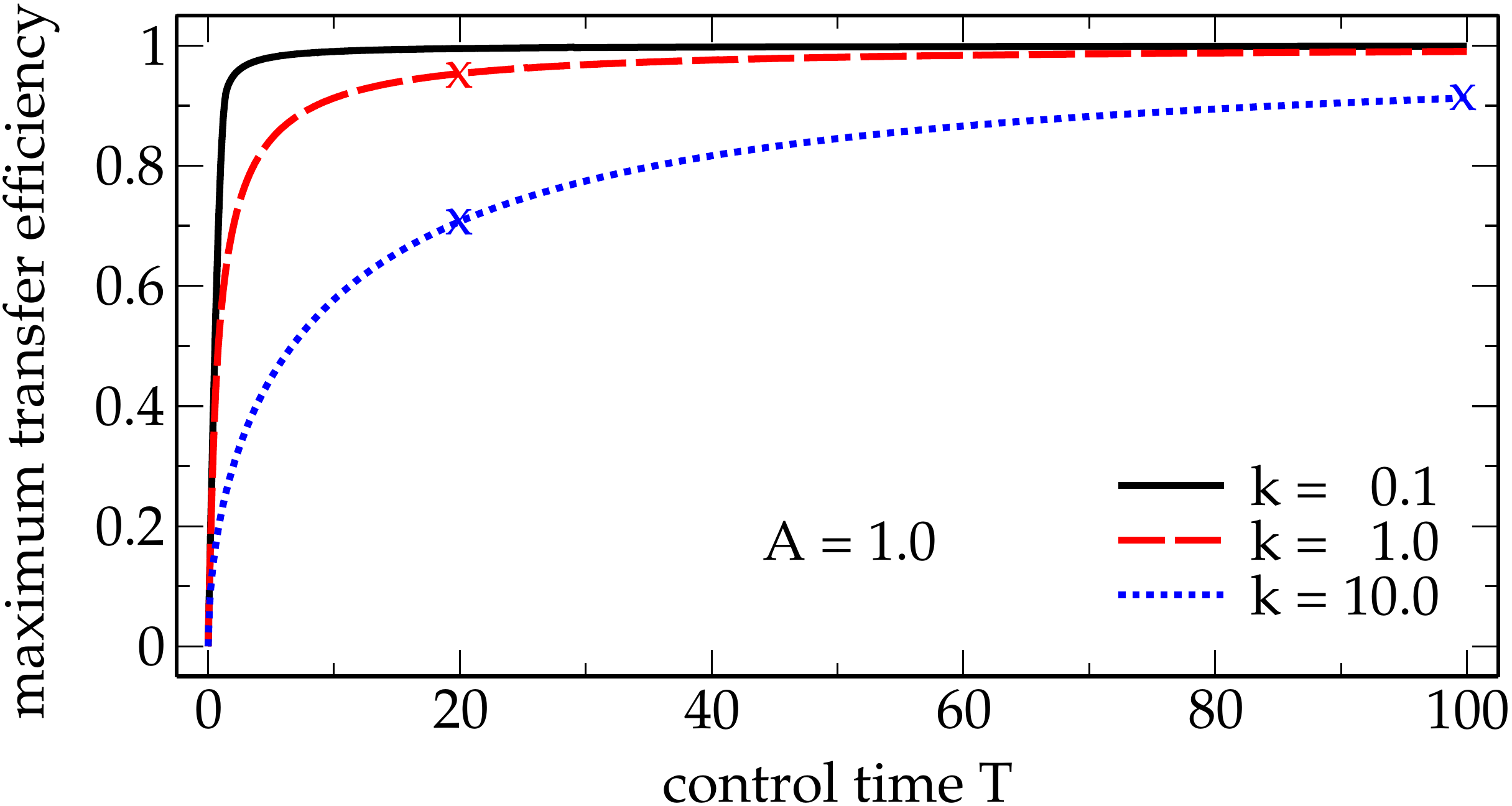}
\end{center}
\caption{(Color online)
  The optimal transfer efficiency, $r_2(T)$ as a function of
  the control time $T$ for different relaxation rates $k$ (the maximum
  Rabi frequency, or pulse power, respectively, is limited by $A=1$).
  The curves are
  fairly concave, reaching a high transfer efficiency in relatively
  short time. This is compatible with the fact that in real
  experiments stimulated Raman adiabatic passage is actually done in
  relatively short time.
  To reach 100\% transfer efficiency, however, infinite
  time is needed.
}
   \label{fig:transfer}
\end{figure}
Substituting the optimal control
$u^*(t)$ into the dynamics, Eq.(\ref{eq:r}), and integrating it yields
the value of $r_2(T)$. For  finite $T$ this gives an upper bound for
the maximum achievable population transfer
due to the possibility of $\Omega_p(t)$ going to infinity.
As shown in Fig.~\ref{fig:transfer} the upper bound approaches unity
in the
limit $T\to\infty$ even for large decay rates $k$, illustrating the
recovery of STIRAP. For small and
moderate decay rates, the convergence toward unit efficiency is much
faster, reflecting the easier control problem.


\end{document}